\documentclass[preprint,superscriptaddress,aps]{revtex4-2}

\usepackage{amsmath, amssymb, graphicx, braket, pgf, hyperref, enumitem}

\usepackage[group-separator={,}]{siunitx}
\DeclareMathOperator{\arccosh}{arccosh}

\let\oldrightarrow\rightarrow
\renewcommand{\rightarrow}{\mathbin{\oldrightarrow}}
\let\oldleftarrow\leftarrow
\renewcommand{\leftarrow}{\mathbin{\oldleftarrow}}
\let\oldleftrightarrow\leftrightarrow
\renewcommand{\leftrightarrow}{\mathbin{\oldleftrightarrow}}
\begin{document}

\title{High-purity entanglement mediated by magnons despite weak coupling}
\author{Sanchar Sharma}
\affiliation{Departamento de Física Teórica de la Materia Condensada, Universidad Autónoma de Madrid, E-28049 Madrid, Spain}
\affiliation{Condensed Matter Physics Center (IFIMAC), Universidad Autónoma de Madrid, E-28049 Madrid, Spain}

\begin{abstract}
Entangling distant spins via a shared magnonic bus typically faces a tradeoff: stronger spin-magnon coupling increases the entanglement fidelity, but also the spin decay rate. We propose a protocol that breaks this tradeoff. The magnons couple only to a transition outside the computational basis, in which the Bell state is created. Our protocol is probabilistic, and weak coupling mainly reduces the success probability, with only a small effect on the fidelity. We analyze a setup where the spins are two nitrogen-vacancy (NV) centers near a magnetic wire. In the absence of NV dephasing, the protocol reaches unit fidelity with a maximally entangled state, for arbitrarily weak coupling. In our simulations via the Monte-Carlo wavefunction approach, the NV-magnon coupling is taken to be one-third of the magnon linewidth. Considering finite NV dephasing, we predict a fidelity of $>0.99$ for a state-of-the-art rate, and an optimal fidelity of $0.91$ for a moderate rate, both at $0.6\%$ success probability.
\end{abstract}

\maketitle

Ferromagnets are well-established in classical applications due to their low losses and scalability~\cite{Roadmap}, and have recently attracted interest for quantum applications~\cite{yuan_quantum_2022}, with several experimental~\cite{QMag_Tabuchi, QMag_Quirion, Xu_MagFock, Xu_Bell} and theoretical~\cite{MagEnt_Mehrdad, MagHer_Victor, ArbitStGen_Mine, QMag_Marios, Coherence_Mehrdad, QTom_Mag, Skyr_Mine} advances. In particular, magnons, the elementary excitations of the magnetization, are promising for quantum transduction~\cite{Martijn_QGates, Fabian_Transduction} because they couple to a wide range of information carriers, such as optical photons~\cite{OptMag_Osada, OptMag_Zhang, OptMag_Silvia, OptMag_James,Fran_coherent_pumping,OMag_WG_Zhu}, microwaves~\cite{MagMW_Th, MagMW_Exp1, MagMW_Exp2}, phonons~\cite{MagPh1, MagPh2}, and nitrogen-vacancy (NV) centers in diamond~\cite{MagNV_Toeno, MagNV_Bertelli, MagNV_Carlos}. Additionally, magnons have coherence lengths of tens of micrometers~\cite{MagNV_Bertelli}, making them attractive as carriers of quantum information.

For a quantum channel transporting information, entanglement distribution between distant qubits is a fundamental task, spanning applications from quantum networks to modular quantum processors~\cite{SC_QC_Rev,QuIC_Rev}. Several entangling protocols using magnons have been proposed~\cite{Loss_Entanglement,Denis_Entanglement,Martijn_Entanglement}, primarily using NV centers as spins. NV centers are of independent interest as quantum memories, because they have extremely long lifetimes, up to seconds, and a dephasing time that dynamical decoupling can extend close to a second~\cite{NV_Review,NV_one_second}. Furthermore, the techniques to initialize NV centers optically and to apply gates using microwave inputs are well-established. Scaling up an NV-based platform requires entangling NV centers on demand, even when they are isolated from each other.

However, these proposals face a major theoretical challenge. Generating sizable entanglement generally requires the NV-magnon coupling to exceed the magnon dissipation rate, which by itself is experimentally challenging. There is also a more fundamental issue. A higher coupling leads to a strong decay of the NVs into the magnon bath. This decay is governed by the same NV-magnon coupling that mediates entanglement, so the tradeoff seems unavoidable at first glance. For example, detuning the NVs can reduce the decay into the magnons, but it also reduces the coupling~\cite{Denis_Entanglement}.

\begin{figure}
    \centering
    \includegraphics[width=0.7\columnwidth]{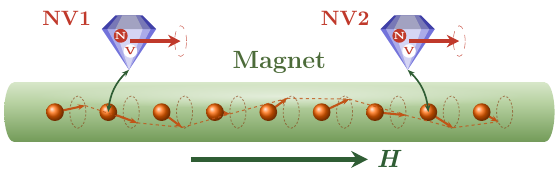}
    \caption{Setup for heralding entanglement between two NV centers, NV1 and NV2. They sit near a thin magnetic wire that mediates the NV-NV entanglement. A static field $\mathbf{H}$ along the wire sets both the equilibrium magnetization and the NV quantization axis. Each NV couples to the magnons via dipolar interactions (double-headed arrows).} \label{fig:setup}
\end{figure}

Our protocol avoids this tradeoff entirely, using an idea similar to the Barrett-Kok (BK) protocol~\cite{Barrett_Kok} in quantum optics. The BK protocol was implemented experimentally in loophole-free tests of Bell-inequality violation~\cite{Loophole_free_Bell}. Adapting the BK protocol to magnons requires solving problems absent in optics. The optical scheme places each NV center in its own cavity, radiating into independent modes that then interfere via a beamsplitter. In our case, see Fig.~\ref{fig:setup}, the two NVs instead share a single magnonic bus, and we use collective decoherence as a resource for building entanglement. As the NVs are expected to be within micrometers, our protocol assumes only global microwave control, with no individual addressing. Such microwaves can also excite the zero-momentum (Kittel) mode of the magnet, but our protocol relies only on magnons with non-zero wavevectors.

The main idea to avoid the aforementioned tradeoff is to couple the magnons to a transition outside the computational basis. Each NV is a spin-1 system with the eigenstates of $S_z$ being $\{\ket{0},\ket{\bar{1}},\ket{1}\}$, with eigenvalues $\{0,-1,1\}$ respectively. We want to create the `desired' Bell state $\ket{\psi_{\rm des}} \propto \ket{0\bar{1}} + \ket{\bar{1}0}$, where $\ket{xy}$ is the state with NV1 in $\ket{x}$ and NV2 in $\ket{y}$. If the magnons are coupled only to the $\ket{0}\leftrightarrow\ket{1}$ transition of each NV, there is no decay inside the computational basis $\{\ket{0}, \ket{\bar{1}}\}$, which avoids the tradeoff. Below, we consider the case when the $\ket{0}\leftrightarrow\ket{\bar{1}}$ transition lies inside the magnon gap. Such a decoupling can also be achieved by magnon chirality in certain geometries~\cite{Martijn_Chiral}. We also assume temperatures to be much smaller than the $\ket{0}\leftrightarrow\ket{1}$ transition frequency.

An NV can flip $\ket{1}\rightarrow\ket{0}$ by emitting a magnon, or $\ket{0}\rightarrow\ket{1}$ by absorbing one. To create entanglement, we want to erase ``which-way'' information, i.e. if a magnon is measured, we do not know which NV emitted it. Thus, this protocol requires a detector of single magnons at non-zero wavevectors. Below, we discuss possibilities to achieve such measurements.

The protocol proceeds as follows:
\begin{enumerate}[itemsep=0pt, parsep=0pt, topsep=0pt, partopsep=0pt]
    \item Prepare each NV in the superposition $\ket{\bar{1}} + \ket{1}$.
    \item \emph{Conditioning click}: Wait for a fixed time $t_1$, continuously monitoring for a magnon. If there is no magnon measured, discard the run.
    \item At $t_1$, apply the cyclic gate $\mathcal{G} \equiv \ket{0}\rightarrow\ket{\bar{1}}\rightarrow\ket{1}\rightarrow\ket{0}$ to both NVs.
    \item \emph{Heralding}: Wait for another fixed time $t_2$, monitoring the magnons. If a magnon is measured, the run is a success.
\end{enumerate}

To understand why this protocol produces a Bell state, $\ket{\psi_{\rm des}}$, we use the Monte-Carlo wavefunction approach, see Appendix~\ref{sec:MCWFmethod}, which is a stochastic unravelling of the Lindblad master equation~\cite{QTraj_OG}. The Hamiltonian $H$ is a coherent exchange, where an NV flips between $\ket{1}$ and $\ket{0}$ by emitting or absorbing a magnon. The dissipation is described by collapse operators $L$ of three kinds: magnon loss into the magnet, magnon measurement, and NV dephasing. NV \emph{decay}, with lifetimes of seconds, is safely neglected. Pure magnon dephasing leaves the fidelity unchanged, see Appendix~\ref{sec:magdeph}. It is ignored here. In the trajectory picture, the joint wavefunction $\ket{\psi}$ is a stochastic variable. It evolves under a non-Hermitian Hamiltonian $\tilde{H} = H - \frac{i}{2}\sum_L L^\dagger L$, interrupted by random collapses $\ket{\psi}\rightarrow L\ket{\psi}$. The explicit forms of $H$ and $L$, along with the detailed trajectories, are given in Appendix~\ref{sec:model}. For now, we only discuss trajectories where no NV dephasing occurred.

Step 1 requires an initialization of both NVs into $\ket{0}$, followed by global microwave pulses creating the superposition. Initialization of an NV is a well-established procedure, see Ref.~\cite{NV_Review} for details. Here, we have assumed that the two NV frequencies are the same, which holds because they sit in the same magnetic field.

If the conditioning click occurs at some time $t_m < t_1$, the wavefunction undergoes an evolution $\ket{\psi} \rightarrow e^{-i\tilde{H}t_m}\ket{\psi}$ followed by a quantum jump such that $\ket{1}\rightarrow\ket{0}$ in one of the NVs. As the information about which NV emitted is not measured, a magnon measurement creates an entangled state (Appendix~\ref{sec:quant})
\begin{equation}
    \ket{\psi} \propto a \ket{B(\theta)} + b \ket{D(\varphi)} + \sum_{\alpha} c_{\alpha} \ket{00\alpha}. \label{psi:step2}
\end{equation}
Here, $\ket{D(\varphi)} = \ket{0\bar{1}} + e^{i\varphi}\ket{\bar{1}0}$ is a dark state as it is not coupled to the magnons. $\ket{B(\theta)} = \ket{01} + e^{i\theta}\ket{10}$ and $\ket{00\alpha}$ each can emit magnons. The coefficients $\{a, b, c_{\alpha}\}$ depend on $t_m$ because of the evolution under $\tilde{H}$. Note that the same state is generated when a magnon instead dissipates into the magnet, which is effectively a measurement by the environment. The experimenter has no way to know that this happened, so externally it falls under ``no conditioning click''.

Two things can happen between $t_m$ and $t_1$. A second magnon can be measured or lost, leaving $\ket{00}$, which has no further decay into the magnons. Otherwise, the bright components decay under $\tilde{H}$ while the dark component does not. Physically, this decay is a Bayesian update. Only the bright components can emit, so the absence of a measurement is evidence that the wavefunction is dark. If $t_1 - t_m$ is much larger than the decay times of the bright components, we end up with only the dark state. As we discuss below, this is the condition under which unit fidelity is achieved.

The dark doublet $\ket{D(\varphi)}$ is already a Bell state, although with an arbitrary phase $\varphi$. At this stage, we cannot tell whether the state is $\ket{00}$ or $\ket{D(\varphi)}$.

To remove the residual bright doublet (for a finite $t_1$) and the trajectories leading to $\ket{00}$, we go to the next steps. We apply the cyclic gate $\mathcal{G}$ to both NVs. This gate can be applied by a $\pi$-pulse on $\ket{0}\leftrightarrow\ket{\bar{1}}$ followed by one on $\ket{0}\leftrightarrow\ket{1}$. If the wavefunction before the gate were $\ket{00}$, it would become $\ket{\bar{1}\bar{1}}$. If it were instead of the form of Eq.~(\ref{psi:step2}) (i.e. only one magnon decay occurred before $t_1$), we get a swap between bright and dark doublets,
\begin{equation}
    \ket{\psi} \propto a' \ket{D(\theta')} + b\left(\ket{1\bar{1}} + e^{i\varphi'}\ket{\bar{1}1}\right) + \sum_{\alpha} c_{\alpha}' \ket{\bar{1}\bar{1}\alpha}, \label{psi:post_gate}
\end{equation}
where $\{\theta', \varphi', a', c_{\alpha}'\}$ are coefficients after evolution until $t_1$. The coefficient of the dark state $b$ does not reduce under $\tilde{H}$. 

We now wait up to a time $t_2$ for another magnon measurement. The $\ket{\bar{1}\bar{1}}$ state cannot emit a magnon, so a measurement implies that we were in Eq.~(\ref{psi:post_gate}) at $t_1$. The second emission can come from the second term transitioning $\ket{1}\rightarrow\ket{0}$ (becoming a dark state henceforth) or the third term emitting a magnon. We are still left with a superposition,
\begin{equation}
    \ket{\psi} \propto b' \ket{D(\varphi'')} + c'' \ket{\bar{1}\bar{1}},
\end{equation}
with reduced coefficients $\{b', c''\}$ and a new phase $\varphi''$, depending on both measurement times. For a sufficiently large $t_1$, the last term is negligible, and unit fidelity is achieved.

The stochastic ensemble of wavefunctions is equivalent to the density matrix $\rho \propto \sum \ket{\psi}\bra{\psi}$ summing over all trajectories. If the phase $\varphi''$ of the component $\ket{D(\varphi'')}$ varied from run to run, the averaging would give a classical density matrix $\rho \propto \ket{0\bar{1}}\bra{0\bar{1}} + \ket{\bar{1}0}\bra{\bar{1}0}$. Appendix~\ref{sec:quant} gives a more explicit form (ignoring the $\ket{\bar{1}\bar{1}}$ component),
\begin{equation}
    \sum_{\alpha\beta} B_{\alpha} B_{\beta} \left(\ket{0\bar{1}} + e^{-i\left(k_{\alpha} - k_{\beta}\right)x}\ket{\bar{1}0}\right), \label{eq:fin_state}
\end{equation}
where the summation runs over magnon modes, the coefficients $B_{\alpha}$ depend on the measurement times, and $x$ is the distance between the two NVs. The phase difference is given by the wavevectors of the magnons $\{k_{\alpha}, k_{\beta}\}$. Now, the NVs couple to a very narrow set of wavevectors, those for which the magnons are resonant, $k_{\alpha} \approx \pm k_{\rm mag}$. In this case, the phase difference is either zero, when $k_{\alpha} \approx k_{\beta}$, or $\pm 2k_{\rm mag}x$, when $k_{\alpha} \approx -k_{\beta}$. Imposing $2k_{\rm mag}x = 2n\pi$ makes every run converge to the same state, $\ket{0\bar{1}} + \ket{\bar{1}0}$. This puts a restriction on the NV frequency and position, which can be achieved by tuning the frequency via the applied magnetic field. 

\begin{figure*}
    \centering
    \includegraphics[width=\textwidth]{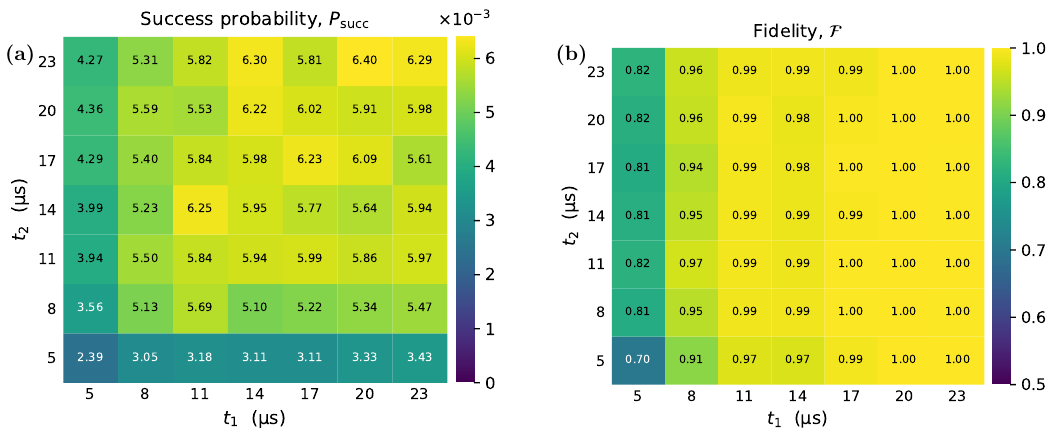}
    \caption{(a) Success probability and (b) fidelity as a function of the two waiting times $t_1$ and $t_2$, for the low NV dephasing rate $\kappa_{\rm NV}^{\rm low}$ (coherence time $\SI{160}{\milli\second}$).}
    \label{fig:prob_fid_low}
\end{figure*}

The derivation so far ignored NV dephasing, which originates from impurities inside the diamond and is unrelated to the magnons. This causes a loss of fidelity, because a trajectory may end up in $\ket{0\bar{1}}-\ket{\bar{1}0}$ instead of the desired state. We evaluate its effect in the numerical simulations, again using the Monte-Carlo wavefunction approach. We consider only the heralded trajectories, i.e. a conditioning click before $t_1$, and a heralding click within the following time $t_2$. The ratio of successful to total trajectories gives the probability of success, $P_{\rm succ}$. To calculate the fidelity, we find the heralded density matrix $\rho_{\rm traj} \propto \sum \ket{\psi}\bra{\psi}$, and the fidelity is $\mathcal{F} = \braket{\psi_{\rm des}|\rho_{\rm traj}|\psi_{\rm des}}$.

Consider a \SI{10}{\micro\meter}-long magnet made of Yttrium Iron Garnet (YIG) with a \SI{40}{\nano\meter} radius, taken to be a cylinder for simplicity. The two NVs are separated by \SI{1}{\micro\meter} and located \SI{5}{\nano\meter} from the magnet surface. For the magnons, we consider the exchange-dipolar dispersion~\cite{StanPrabh,Kalinikos_Slavin} assuming a uniform mode across the cross-section and a wavevector along the wire axis parallel to the magnetization. This dispersion is an estimate for a wire with unpinned surface spins, see Appendix~\ref{sec:dispersion} for further discussion. The estimate works for our case because the only parameter that we need from it is the wavevector of magnons resonant with the NVs, which in turn only has a weak effect on the coupling. A magnetic field of $\sim \SI{70}{\milli\tesla}$ is chosen to ensure the wavevector resonance discussed above. Then, the $\ket{0}\leftrightarrow\ket{1}$ transition is at $2\pi\times\SI{4.83}{\giga\hertz}$, resonant with the two wavevectors $\pm k_{\rm mag}$, where $2\pi/k_{\rm mag} \approx \SI{200}{\nano\meter}$. The same field places the $\ket{0}\leftrightarrow\ket{\bar{1}}$ transition at $2\pi\times\SI{0.9}{\giga\hertz}$, below the lowest magnon frequency of $2\pi\times\SI{3.68}{\giga\hertz}$. The resulting coupling is $g = 2\pi\times\SI{30}{\kilo\hertz}$, about three times smaller than the assumed magnon linewidth $\kappa = 2\pi\times \SI{0.1}{\mega\hertz}$, corresponding to a lifetime of \SI{1.7}{\micro\second}. This is conservative compared to the measured \SI{18}{\micro\second} of high-momentum magnons~\cite{Rostyslav_longlivingmagnons}.

We now turn to the NV dephasing rate. Dynamical decoupling can push the NV coherence times towards the order of seconds~\cite{NV_one_second}, far longer than the microsecond timescales $t_1,t_2$ of the protocol, so the dephasing accumulated over a run can be small. The decoupling pulses are global microwave pulses as well. A rigorous treatment of dynamical decoupling is beyond the scope of this work. Qualitatively, such pulses imply that the NVs spend only half their time in the excited state, $\ket{1}$, reducing the effective coupling by a factor of $2$. This can be mitigated, e.g., by a smaller magnetic wire. Below, we study two NV dephasing rates. We first take a small dephasing rate $\kappa_{\rm NV}^{\rm low} = 2\pi\times\SI{1}{\hertz} = 1/\SI{160}{\milli\second}$ reachable with dynamical decoupling~\cite{NV_one_second}. We then analyze the case of a moderate rate $\kappa_{\rm NV}^{\rm mod} = 2\pi\times\SI{1}{\kilo\hertz} = 1/\SI{160}{\micro\second}$.

For a magnon to be measured rather than lost into the magnet, the measurement rate must be at least a sizable fraction of the magnon decay rate. In the simulations below, we assume that only the $+k_{\rm mag}$ magnons are measured, at a rate of $\Gamma = 2\pi\times\SI{30}{\kilo\hertz}$. This equals the coupling $g$ and is about three times below the linewidth of each magnon $\kappa$. Counting magnons with negative wavevector, this amounts to a \emph{6-fold smaller} rate of measurement compared to decay into the magnet.

Fig.~\ref{fig:prob_fid_low} shows the success probability and fidelity as functions of $t_1$ and $t_2$ at the low dephasing rate $\kappa_{\rm NV}^{\rm low}$, each obtained by averaging $\num{100000}$ quantum trajectories. The probability increases with both waiting times, since longer windows give higher chances of measuring the heralding magnons. It saturates near $0.6\%$ once both times exceed roughly \SI{10}{\micro\second}. The fidelity increases monotonically with $t_1$ and is nearly independent of $t_2$. The infidelity at small $t_1$ comes from the residual magnon of the $\ket{00\alpha}$ term in Eq.~(\ref{psi:step2}), as discussed above. Dephasing at this rate is negligible over the protocol duration, so the fidelity exceeds $0.99$ for $t_1\ge\SI{17}{\micro\second}$.

\begin{figure*}
    \centering
    \includegraphics[width=\textwidth]{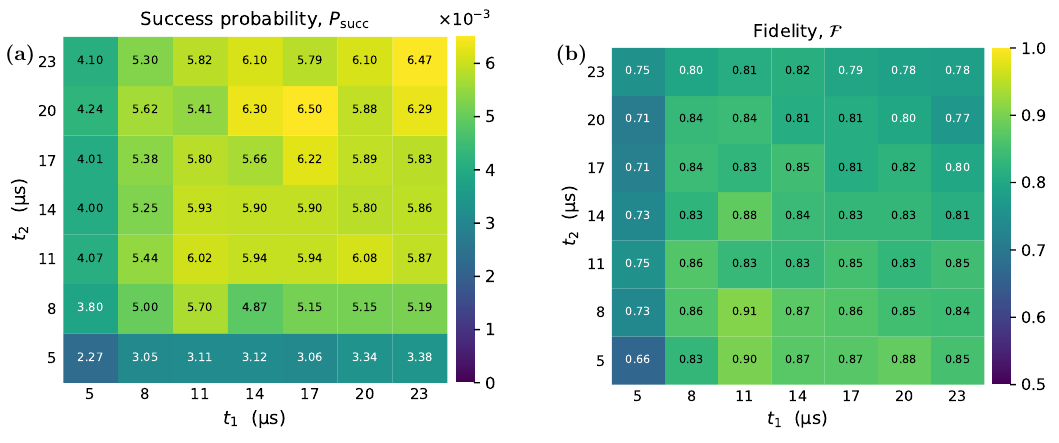}
    \caption{(a) Success probability and (b) fidelity as a function of the two waiting times $t_1$ and $t_2$, for the moderate NV dephasing rate $\kappa_{\rm NV}^{\rm mod}$ (coherence time $\SI{160}{\micro\second}$).}
    \label{fig:prob_fid_high}
\end{figure*}

We now turn to the moderate rate $\kappa_{\rm NV}^{\rm mod}$, shown in Fig.~\ref{fig:prob_fid_high}. The success probability is nearly unchanged, as dephasing does not alter the emission of magnons. The fidelity, however, now peaks at an intermediate $t_1\approx\SI{10}{\micro\second}$, because at large $t_1$ the NV dephasing starts to take over. The fidelity also decreases with $t_2$ for the same reason. The optimum is $\mathcal{F} = 0.91$ at $t_1 = \SI{11}{\micro\second}$ and $t_2 = \SI{8}{\micro\second}$, with a success probability of $0.57\%$. The waiting times thus face competing requirements. To ensure a reasonable success probability, both $t_1$ and $t_2$ should be at least comparable to the inverse measurement frequency. However, they must be much shorter than the NV dephasing time to avoid phase slips. An experiment therefore requires NV dephasing to be much slower than magnon decay and measurement.

The success probability translates into an entanglement rate. The rate of generating Bell pairs is $P_{\rm succ}/\left(t_1 + P_{\rm first} t_2 + t_{\rm over}\right)$, where $P_{\rm first}$ is the probability of a magnon measurement before $t_1$, and the overhead $t_{\rm over} = \SI{5}{\micro\second}$~\cite{NV_Network} covers the NV initialization and gates. At the optimal fidelity point noted above, this rate is $\SI{0.35}{\kilo\hertz}$. Reducing the coupling to $g = 2\pi\times\SI{20}{\kilo\hertz}$ moves the optimum to $t_1 = \SI{17}{\micro\second}$ and $t_2 = \SI{11}{\micro\second}$, where $\mathcal{F} = 0.88$ and $P_{\rm succ} = 0.43\%$. A weaker coupling thus costs entanglement rate, which falls to $\SI{0.18}{\kilo\hertz}$, and has only a weak effect on the fidelity.

We discuss the effects of different experimental non-idealities qualitatively. A detuning $\delta$ between the two NVs, potentially due to inhomogeneities, can make the initializations and gates challenging to implement. It also makes the two emission paths accumulate different phases over the random click times, which requires $\delta \ll 1/(t_1 + t_2) \sim 2\pi\times\SI{10}{\kilo\hertz}$. Stochastic variations in the magnetic field shift the resonance condition $k_{\rm mag} x = n\pi$, by varying $k_{\rm mag}$. The resulting fidelity loss is quadratic in the wavevector error. The same phase-matching condition also constrains the NV-NV distance, which would be fixed at fabrication. The applied field should then be tuned to satisfy the resonance condition. In our case, for an arbitrary $x$ within $\SI{1}{\micro\meter} \pm \SI{20}{\nano\meter}$, we can always find a magnetic field within $\numrange{50}{200}\,\si{\milli\tesla}$ to satisfy the resonance. Finally, we assume a temperature well below the $\ket{0}\leftrightarrow\ket{1}$ transition, $T \ll \SI{230}{\milli\kelvin}$.

The most experimentally challenging component of this protocol is single-magnon detection at high wavevectors. Note that the protocol requires a sensitive detector, but it need not be coherent. Such a detection has been demonstrated for the Kittel mode, using a superconducting qubit~\cite{QMag_Quirion,lachance-quirion_entanglement-based_2020,Xu_MagFock}. Microwave coupling at non-zero wavevectors is typically zero, but a grating can supply the missing momentum~\cite{BLS_NanoAntenna}. Beyond a qubit, Brillouin light scattering (BLS) or another NV center can also be used for detecting magnons, however their sensitivity is typically low. Thus a magnon amplifier is required. We discuss an amplification using the three-magnon splitting~\cite{NL_Rev, Suhl_OG}, already analyzed theoretically in the classical regime~\cite{Xu_3Mag}. The magnon dispersion admits a resonant triplet: a pump mode at $k_{\rm pump} = \SI{66.65}{\per\micro\meter}$ ($\omega_{\rm pump} = 2\pi\times\SI{10.03}{\giga\hertz}$) splits into the NV-coupled mode at $k_{\rm mag} = \SI{31.41}{\per\micro\meter}$ ($\omega_{\rm mag} = 2\pi\times\SI{4.83}{\giga\hertz}$) and an idler mode at $k_{\rm idler} = \SI{35.24}{\per\micro\meter}$ ($\omega_{\rm idler} = 2\pi\times\SI{5.20}{\giga\hertz}$). These satisfy $\omega_{\rm pump} = \omega_{\rm mag} + \omega_{\rm idler}$ and $k_{\rm pump} = k_{\rm mag} + k_{\rm idler}$. A grating coupler can generate the pump locally, away from the NV centers. 

Quantum mechanically, the process is governed by a parametric interaction term, $\propto g_{\rm eff} \hat{m}_{\rm mag}^{\dagger} \hat{m}_{\rm idler}^{\dagger}$. Under strong coupling, $g_{\rm eff} > \{\kappa_{\rm mag}, \kappa_{\rm idler}\}$, and assuming equal linewidths $\kappa_{\rm mag} = \kappa_{\rm idler}$ for simplicity, it gives an amplification of the form,
\begin{equation}
    \hat{m}_{\rm mag}(t) = \cosh(g_{\rm eff} t) \hat{m}_{\rm mag}(0) + \sinh(g_{\rm eff} t) \hat{m}^{\dagger}_{\rm idler}(0).
\end{equation}
The second term is noise coming from the idler magnons. Starting from vacuum, $n_{\rm mag}(0) = n_{\rm idler}(0) = 0$, the output is $n_{\rm mag}(t) = \sinh^2(g_{\rm eff} t)$. A single magnon, $n_{\rm mag}(0) = 1$, raises it to $n_{\rm mag}(t) = \cosh^2(g_{\rm eff} t) + \sinh^2(g_{\rm eff} t)$, with a large contrast $\cosh^2(g_{\rm eff} t)$ and a moderate signal-to-noise ratio below $2$. For $g_{\rm eff} = 2\kappa_{\rm mag} = 2\pi\times\SI{0.2}{\mega\hertz}$, and assuming that we need an amplification up to $1000$ magnons (the room-temperature limit), the measurement rate is $\Gamma \sim g_{\rm eff} / \arccosh(\sqrt{1000}) = 2\pi\times\SI{48}{\kilo\hertz}$. This exceeds the rate $2\pi\times\SI{30}{\kilo\hertz}$ assumed in the simulations.

Note that this mechanism does not directly fit the constant measurement rate assumed in our simulations, because here the measurement rate increases with time. A thorough analysis including this effect is beyond the scope of this work.

In summary, we proposed a heralded protocol to entangle two NV centers through a shared magnonic bus, breaking the coupling-decoherence tradeoff present in existing schemes. The magnons couple only to a transition outside the computational basis, so the computational states never see the magnon bath. A weak coupling then mainly reduces the probability of success, and has only a small effect on the fidelity. In our example, the coupling is three times below the magnon linewidth, and yet we predict fidelities $\mathcal{F} = 0.91$ at an NV dephasing rate of $2\pi\times\SI{1}{\kilo\hertz}$, and $\mathcal{F} > 0.99$ at $2\pi\times\SI{1}{\hertz}$, both at a success probability near $0.6\%$. The most experimentally challenging component is the single-magnon detection at nonzero wavevector, for which we proposed parametric amplification by three-magnon splitting. The protocol does not require individual control of the NVs, making it suitable for on-chip experiments. As the main mechanism of entanglement is the erasure of the information about ``which NV emitted a magnon'', this protocol could be scalable to more NVs, creating high-dimensional entangled states.

\begin{acknowledgments}
The project that gave rise to these results received the support of a fellowship from ``la Caixa'' Foundation (ID 100010434). The fellowship code is LCF/BQ/PI25/12100023. This publication is part of the grant ``Excelencia María de Maeztu'' CEX2023-001316-M funded by MICIU/AEI /10.13039/501100011033. The figures in this manuscript can be reconstructed with the openly available code~\cite{NV_Magnon_NV_Code}.
\end{acknowledgments}

\clearpage

\section*{Appendix}

There are two appendices. Appendix~\ref{sec:MCWF} analyzes the heralded protocol with the Monte-Carlo wavefunction method. It tracks the state through the two magnon measurements and shows that a successful herald leaves the two NVs in the target Bell state. Appendix~\ref{sec:modes} describes the magnon dispersion of the wire, the stray field of a single magnon, and the resulting coupling to the NV centers. It fixes the numbers quoted in the main text.

\appendix

\section{Analysis via Monte-Carlo wavefunctions} \label{sec:MCWF}
In this appendix, we analyze the quantum trajectories that emerge from the Monte-Carlo approach.

\subsection{Model} \label{sec:model}

The Hamiltonian is
\begin{equation}
    \frac{H_{\rm main}}{\hbar} = \sum_j \left(\omega_1^j P_1^j + \omega_{\bar{1}}^j P_{\bar{1}}^j\right)
    + \sum_\alpha \omega_\alpha m_\alpha^\dagger m_\alpha
    + \sum_{\alpha j} \left(g_\alpha e^{ik_\alpha x_j} m_\alpha P_{01}^j + g_\alpha^* e^{-ik_\alpha x_j} m_\alpha^\dagger P_{10}^j\right). \label{Seq:ham}
\end{equation}
We divide every Hamiltonian by $\hbar$, so that $\omega_\alpha$, $g_\alpha$, and the eigenvalues below are angular frequencies. Here, $P_{ab}^j = \ket{b}_j \! \bra{a}$ is the transition operator $\ket{a} \rightarrow \ket{b}$ of the NV labeled $j \in \{l,r\}$ (left and right), located at $x_j$ along the wire, with $P_a^j \equiv P_{aa}^j$ the projector onto $\ket{a}$. The frequencies $\omega_1^j$ and $\omega_{\bar{1}}^j$ are those of the $\ket{0} \leftrightarrow \ket{1}$ and $\ket{0} \leftrightarrow \ket{\bar{1}}$ transitions. The operator $m_\alpha$ annihilates a magnon of mode $\alpha$ with wavevector $k_\alpha$ and frequency $\omega_\alpha$. The coupling amplitude $g_\alpha$ is derived in Appendix~\ref{sec:modes}. An NV emitting a magnon into mode $\alpha$ picks up the position-dependent phase $e^{-ik_\alpha x_j}$, which controls the phase of the final Bell state.

There are three collapse channels, described by
\begin{equation}
    L \in \left\{ \sqrt{\kappa_\alpha} \, m_\alpha, \; \sqrt{\Gamma} \, \hat{M}, \; \sqrt{\kappa_{\rm NV} / 4} \, \left(I - 2 P_i^j\right) \right\} .
\end{equation}
The first is magnon loss, with $\kappa_\alpha$ the dissipation rate of magnon mode $\alpha$. The second is the magnon measurement, at the rate $\Gamma$. The measurement operator is a linear combination of magnon modes,
\begin{equation}
    \hat{M} = \sum_\alpha O_\alpha m_\alpha, \qquad \sum_\alpha \left|O_\alpha\right|^2 = 1, \label{Seq:meas_op}
\end{equation}
with the coefficients $O_\alpha$ determined by the measurement apparatus. The third is NV dephasing, one operator for each NV $j$ and state $i$. This collapse flips the sign of the $\ket{i}$ component. It exchanges no energy and creates no magnons. The factor of $4$ ensures that the off-diagonal entries of the density matrix decay at the rate $\kappa_{\rm NV}$.

We retain NV dephasing in the non-Hermitian Hamiltonian, but we do not track the trajectories in which a dephasing jump occurs. Since $(I - 2 P_i^j)^2 = I$, these collapse operators add a term proportional to the identity to $\widetilde{H}$. NV dephasing thus enters as a uniform decay of every state, set by the total dephasing rate $\kappa_{\rm NV}$, independent of the NV configuration. This uniform decay cancels once the trajectory is normalized.

For the joint states, we write $\ket{ab}$ for NV$_l$ in $\ket{a}$ and NV$_r$ in $\ket{b}$ with the magnons in vacuum. A trailing Greek index adds a magnon, $\ket{ab\alpha} = m_\alpha^\dagger \ket{ab}$. For two magnons, $\ket{ab\alpha\beta}$ denotes the normalized state, i.e., $\ket{ab\alpha\beta} = m_\alpha^\dagger m_\beta^\dagger \ket{ab}$ for $\alpha \ne \beta$ and $\ket{ab\alpha\alpha} = (m_\alpha^\dagger)^2 \ket{ab}/\sqrt{2}$.

Below, we go into the rotating frame via the unitary transformation
\begin{equation}
    U(t) = e^{-i(\omega_1^l + \omega_1^r)t/2 \left(\sum_\alpha m_\alpha^\dagger m_\alpha + P_1^l + P_1^r\right)} e^{-it \sum_j \omega_{\bar{1}}^j P_{\bar{1}}^j },
\end{equation}
giving the transformed Hamiltonian
\begin{equation}
    \frac{H}{\hbar} = \Delta \left(P_1^l - P_1^r\right)
    + \sum_\alpha \eta_\alpha m_\alpha^\dagger m_\alpha
    + \sum_{\alpha j} \left(g_\alpha e^{ik_\alpha x_j} m_\alpha P_{01}^j + g_\alpha^* e^{-ik_\alpha x_j} m_\alpha^\dagger P_{10}^j\right),
\end{equation}
where $\Delta = \frac{\omega_1^l - \omega_1^r}{2}$, and $\eta_\alpha = \omega_\alpha - \frac{\omega_1^l + \omega_1^r}{2}$. Below, we assume that both NVs sit in the same static field. Their $\ket{0} \leftrightarrow \ket{1}$ frequencies then match, $\omega_1^l = \omega_1^r$, so that $\Delta = 0$.

\subsection{Monte-Carlo wavefunction method} \label{sec:MCWFmethod}

The Monte-Carlo wavefunction method~\cite{QTraj_OG} unravels the Lindblad master equation into stochastic trajectories of pure states. Each trajectory is a sequence of collapses, one per action of a collapse operator $L$. Between two collapses, the state evolves under the non-Hermitian effective Hamiltonian
\begin{equation}
    \frac{\widetilde{H}}{\hbar} = \frac{H}{\hbar} - \frac{i}{2} \sum_L L^\dagger L, \label{Seq:Heff}
\end{equation}
where the sum runs over all collapse channels. A trajectory is generated by repeating three steps.

First, we propagate the state from the last collapse, at time $t_c$, with the unnormalized evolution
\begin{equation}
    \ket{\tilde{\psi}(t_c + t)} = e^{-i \widetilde{H} t / \hbar} \, \ket{\psi(t_c)} .
\end{equation}
Second, we draw the waiting time $t$ to the next collapse. The squared norm $\braket{\tilde{\psi}(t_c + t) | \tilde{\psi}(t_c + t)}$ decays with $t$, and equals the probability that no collapse has occurred in that interval. It therefore fixes the distribution of $t$, from which we sample. Third, we pick the channel that collapses. Each $L$ is chosen with a probability proportional to $\braket{\tilde{\psi} | L^\dagger L | \tilde{\psi}}$, evaluated at the sampled time. The state right after the collapse is $\ket{\psi(t_c + t)} = \mathcal{N}\left[L \ket{\tilde{\psi}(t_c + t)}\right]$, where $\mathcal{N}$ denotes normalization. The three steps then repeat from this state. At any time between collapses, the physical state is the normalized $\mathcal{N}[\ket{\tilde{\psi}}]$.

\subsection{Subspaces of the non-Hermitian Hamiltonian}

The non-Hermitian Hamiltonian $\widetilde{H}$ conserves the excitation number $\hat{N}_{\rm exc} = P^l_1 + P^r_1 + \sum_\alpha m_\alpha^\dagger m_\alpha$. Its eigenstates therefore split into the independent subspaces listed in the table below. We call a subspace `dark' if no state in it can emit a magnon, and `bright' otherwise. The four subspaces with $\hat{N}_{\rm exc} = 0$ are singletons. Of the two-excitation subspaces, we need only $\mathcal{H}^2_{lr}$. We analyze the protocol through the eigenstates of $\widetilde{H}$ in these subspaces.

\begin{center}
    \begin{ruledtabular}
    \begin{tabular}{ccllc}
        $\hat{N}_{\rm exc}$ & Subspace & Basis & Type & Magnon Decay \\
        \colrule
        $0$ & $\mathcal{H}^0_{ij}$ & $\ket{ij}, \; i,j \in \{0,\bar{1}\}$ & dark & $0$ \\
        \colrule
        $1$ & $\mathcal{H}^1_D$ & $\ket{\bar{1}\bar{1}\alpha}$ & dark & $\mathcal{H}^0_{\bar{1}\bar{1}}$ \\
        $1$ & $\mathcal{H}^1_l$ & $\ket{1\bar{1}}, \; \ket{0\bar{1}\alpha}$ & bright & $\mathcal{H}^0_{0\bar{1}}$ \\
        $1$ & $\mathcal{H}^1_r$ & $\ket{\bar{1}1}, \; \ket{\bar{1}0\alpha}$ & bright & $\mathcal{H}^0_{\bar{1}0}$ \\
        $1$ & $\mathcal{H}^1_{lr}$ & $\ket{10}, \; \ket{01}, \; \ket{00\alpha}$ & bright & $\mathcal{H}^0_{00}$ \\
        \colrule
        $2$ & $\mathcal{H}^2_{lr}$ & $\ket{11}, \; \ket{10\alpha}, \; \ket{01\alpha}, \; \ket{00\alpha\beta}$ & bright & $\mathcal{H}^1_{lr}$ \\
    \end{tabular}
    \end{ruledtabular}
\end{center}
These subspaces are preserved by the non-Hermitian Hamiltonian, and they transform simply under collapses. NV phase flips map each subspace to itself. Magnon decay and measurement both apply $m_\gamma$, or a linear combination thereof. Both therefore send a subspace to the one given in the last column of the table. These maps track the wavefunction through the protocol.

\subsection{Quantum trajectories: subspace analysis}

We now study how the subspace of the wavefunction evolves under the protocol. We expand the initial state in the eigenbasis of $\widetilde{H}$, evolve it under the non-Hermitian Hamiltonian, and apply the collapse operators at the randomly sampled times of clicks.

Step 1 of the protocol prepares each NV in $\ket{\bar{1}} + \ket{1}$ with the magnons in vacuum. Hence $\ket{\psi} \in \mathcal{H}^2_{lr} + \mathcal{H}^1_l + \mathcal{H}^1_r + \mathcal{H}^0_{\bar{1}\bar{1}}$. The protocol rejects any run with no magnon click before the waiting time $t_1$. We therefore keep only the trajectories that clicked.

Three outcomes remain. There were two clicks, one click plus one magnon decay, or a single click. The single-click case gives a wavefunction in $\mathcal{H}^1_{lr} + \mathcal{H}^0_{0\bar{1}} + \mathcal{H}^0_{\bar{1}0}$. The other two give a wavefunction in $\mathcal{H}^0_{00}$. As discussed in the main text, this second outcome eventually leads to failure. We therefore keep only the single-click case.

We now apply the cyclic gate $\mathcal{G} \equiv \ket{0} \rightarrow \ket{\bar{1}} \rightarrow \ket{1} \rightarrow \ket{0}$. This maps $\mathcal{H}^1_{lr} \rightarrow \mathcal{H}^0_{0\bar{1}} + \mathcal{H}^0_{\bar{1}0} + \mathcal{H}^1_D$, where the first two are dark and the third is undesired. The gate also maps $\mathcal{H}^0_{0\bar{1}} \rightarrow \mathcal{H}^1_r$ and $\mathcal{H}^0_{\bar{1}0} \rightarrow \mathcal{H}^1_l$. A second magnon measurement removes the components with no magnons. Acting on $\mathcal{H}^1_l$ and $\mathcal{H}^1_r$, it produces a wavefunction in $\mathcal{H}^0_{0\bar{1}} + \mathcal{H}^0_{\bar{1}0}$, that is, $\ket{\psi} = a \ket{0\bar{1}} + b \ket{\bar{1}0}$. The same measurement also maps $\mathcal{H}^1_D \rightarrow \mathcal{H}^0_{\bar{1}\bar{1}}$, adding a small $\ket{\bar{1}\bar{1}}$ contribution. This term is suppressed at large $t_1$, as discussed in the main text, so we omit it here. The coefficients $a$ and $b$ vary from trajectory to trajectory, and need not form a maximally entangled state. We therefore turn to a more careful analysis.

\subsection{Eigenspace of the single-excitation sectors}
The quantitative analysis below uses the eigenstates of $\mathcal{H}^1_l$ and $\mathcal{H}^1_r$. The matrix elements of $\widetilde{H}$ within the $\mathcal{H}^1_l$ sector, spanned by $\ket{1\bar{1}}$ and $\ket{0\bar{1}\alpha}$, are
\begin{equation}
    \begin{split}
        \frac{\widetilde{H}}{\hbar} \ket{1\bar{1}} &= -i \frac{\kappa_{\rm NV}}{2} \ket{1\bar{1}} + \sum_\alpha g_\alpha^* e^{-ik_\alpha x_l} \ket{0\bar{1}\alpha}, \\
        \frac{\widetilde{H}}{\hbar} \ket{0\bar{1}\alpha} &= \left(\eta_\alpha - i \frac{\kappa_\alpha + \kappa_{\rm NV}}{2}\right) \ket{0\bar{1}\alpha}
        + g_\alpha e^{ik_\alpha x_l} \ket{1\bar{1}}
        - \frac{i \Gamma O_\alpha}{2} \sum_\beta O_\beta^* \ket{0\bar{1}\beta}. \label{Seq:Heff_l}
    \end{split}
\end{equation}
We look for an eigenstate of $\widetilde{H}$ in this sector of the form
\begin{equation}
    \ket{l(\varepsilon)} = \mu_\varepsilon^l \ket{1\bar{1}} + \sum_\alpha \nu_{\varepsilon\alpha}^l \, e^{-ik_\alpha x_l} \ket{0\bar{1}\alpha} .
\end{equation}
Substituting into $\widetilde{H} \ket{l(\varepsilon)} = \hbar \varepsilon \ket{l(\varepsilon)}$ and matching the coefficients of $\ket{1\bar{1}}$ gives
\begin{equation}
    \left(\varepsilon + i \frac{\kappa_{\rm NV}}{2}\right) \mu_\varepsilon^l = \sum_\alpha \nu_{\varepsilon\alpha}^l \, g_\alpha, \label{H1l:eigvec}
\end{equation}
Matching the coefficients of $\ket{0\bar{1}\alpha}$ gives
\begin{equation}
    \left(\varepsilon - \bar{\eta}_\alpha\right) \nu_{\varepsilon\alpha}^l = g_\alpha^* \mu_\varepsilon^l - \frac{i \Gamma}{2} \sum_\beta O_\beta O_\alpha^* \, e^{i(k_\alpha - k_\beta) x_l} \, \nu_{\varepsilon\beta}^l ,
\end{equation}
where $\bar{\eta}_\alpha \equiv \eta_\alpha - i \frac{\kappa_\alpha + \kappa_{\rm NV}}{2}$ collects the complex terms of Eq.~\eqref{Seq:Heff_l}.

The corresponding equations for $\mathcal{H}^1_r$ follow by replacing $x_l$ with $x_r$. The measurement term then carries the phase $e^{i(k_\alpha - k_\beta) x_r}$ instead of $e^{i(k_\alpha - k_\beta) x_l}$. The two sectors therefore share the same eigenvalues when $e^{i(k_\alpha - k_\beta)(x_l - x_r)} = 1$ for all relevant magnons. This condition is easy to meet. The NVs couple only to a narrow set of wavevectors resonant with their frequency, say $\pm k$. We then need only
\begin{equation}
    2k(x_l - x_r) = 2n\pi. \label{H1l:cond}
\end{equation}

We assume this condition holds, and drop the superscript $l$ from the coefficients. The eigenvectors of the two sectors then coincide, up to the phases $e^{-ik_\alpha x_j}$ carried by their magnon components. The eigenstate of the $\mathcal{H}^1_r$ sector at the same eigenvalue $\varepsilon$ is
\begin{equation}
    \ket{r(\varepsilon)} = \mu_\varepsilon \ket{\bar{1}1} + \sum_\alpha \nu_{\varepsilon\alpha} \, e^{-ik_\alpha x_r} \ket{\bar{1}0\alpha} ,
\end{equation}
with the same coefficients $\mu_\varepsilon$ and $\nu_{\varepsilon\alpha}$ as $\ket{l(\varepsilon)}$.

To invert this relation, we need the left eigenvectors $\widetilde{H}^\dagger \ket{L(\varepsilon)} = \hbar \varepsilon^* \ket{L(\varepsilon)}$. Writing $\ket{L(\varepsilon)} = \bar{\mu}_\varepsilon \ket{1\bar{1}} + \sum_\alpha \bar{\nu}_{\varepsilon\alpha} e^{-ik_\alpha x_l} \ket{0\bar{1}\alpha}$, their coefficients satisfy
\begin{equation}
    \left(\varepsilon^* - i \frac{\kappa_{\rm NV}}{2}\right) \bar{\mu}_\varepsilon = \sum_\alpha \bar{\nu}_{\varepsilon\alpha} \, g_\alpha ,
\end{equation}
and
\begin{equation}
    \left(\varepsilon^* - \bar{\eta}_\alpha^*\right) \bar{\nu}_{\varepsilon\alpha} = g_\alpha^* \bar{\mu}_\varepsilon + \frac{i \Gamma}{2} \sum_\beta O_\beta O_\alpha^* \, e^{i(k_\alpha - k_\beta) x_l} \, \bar{\nu}_{\varepsilon\beta} .
\end{equation}
The left eigenvector of the $\mathcal{H}^1_r$ sector, denoted $\ket{R(\varepsilon)}$, follows from $\ket{L(\varepsilon)}$ by the same replacements as above, $\ket{1\bar{1}} \rightarrow \ket{\bar{1}1}$, $\ket{0\bar{1}\alpha} \rightarrow \ket{\bar{1}0\alpha}$, and $x_l \rightarrow x_r$.

We normalize the eigenvectors so that the left and right sets are biorthonormal,
\begin{equation}
    \braket{L(\varepsilon) | l(\varepsilon')} = \braket{R(\varepsilon) | r(\varepsilon')} = \delta_{\varepsilon\varepsilon'} .
\end{equation}
Any state in the $\mathcal{H}^1_l$ sector then expands as $\ket{\chi} = \sum_\varepsilon \braket{L(\varepsilon) | \chi} \ket{l(\varepsilon)}$, and similarly in $\mathcal{H}^1_r$ with $L, l$ replaced by $R, r$.

\subsection{Quantum trajectories: quantitative analysis} \label{sec:quant}

At step 1, the wavefunction is
\begin{equation}
    \text{Step 1: }\ket{\psi} = \frac{\ket{11}+\ket{\bar{1}1}+\ket{1\bar{1}}+\ket{\bar{1}\bar{1}}}{2}. \label{Seq:init}
\end{equation}
By the subspace analysis above, only the middle two terms need to be followed. The last term $\ket{\bar{1}\bar{1}}$ is removed by the first magnon measurement. The first term $\ket{11}$ eventually produces an undesired contribution, suppressed at large $t_1$.

Consider the term $\ket{1\bar{1}}$. We expand it in the eigenbasis of the $\mathcal{H}^1_l$ sector,
\begin{equation}
    \ket{1\bar{1}} = \sum_\varepsilon \braket{L(\varepsilon) | 1\bar{1}} \ket{l(\varepsilon)} = \sum_\varepsilon \bar{\mu}_\varepsilon^* \ket{l(\varepsilon)} .
\end{equation}

Suppose a magnon measurement occurs at $t_m < t_1$. The unnormalized state after the collapse $\sqrt{\Gamma} \, \hat{M}$ is
\begin{equation}
    \left(\sqrt{\Gamma} \sum_{\alpha, \varepsilon} e^{-i\varepsilon t_m - ik_\alpha x_l} \, O_\alpha \bar{\mu}_\varepsilon^* \nu_{\varepsilon\alpha} \right) \ket{0\bar{1}} .
\end{equation}
The state $\ket{0\bar{1}}$ evolves trivially until $t_1$, acquiring a factor $e^{-\kappa_{\rm NV}(t_1 - t_m)/2}$. We then apply the gate to get
\begin{equation}
    \left(\sqrt{\Gamma} \, e^{-\kappa_{\rm NV}(t_1 - t_m)/2} \sum_{\alpha, \varepsilon} e^{-i\varepsilon t_m - ik_\alpha x_l} \, O_\alpha \bar{\mu}_\varepsilon^* \nu_{\varepsilon\alpha}\right) \ket{\bar{1}1} .
\end{equation}
Suppose a second magnon measurement occurs at $t_1 + t_m'$. We evolve until then and collapse. We expand $\ket{\bar{1}1}$ in the eigenbasis of the $\mathcal{H}^1_r$ sector,
\begin{equation}
    \ket{\bar{1}1} = \sum_{\varepsilon'} \braket{R(\varepsilon') | \bar{1}1} \ket{r(\varepsilon')} = \sum_{\varepsilon'} \bar{\mu}_{\varepsilon'}^* \ket{r(\varepsilon')} .
\end{equation}
Here $\ket{r(\varepsilon')}$ and $\ket{R(\varepsilon')}$ are the right and left eigenvectors of the $\mathcal{H}^1_r$ sector, defined above. They share the coefficients $\mu, \nu$ with the $\mathcal{H}^1_l$ sector. The second collapse then gives
\begin{equation}
    \left(\Gamma \, e^{-\kappa_{\rm NV}(t_1 - t_m)/2} \sum_{\alpha, \alpha', \varepsilon, \varepsilon'} e^{-i\varepsilon t_m - i\varepsilon' t_m' - i(k_\alpha x_l + k_{\alpha'} x_r)} \, O_\alpha O_{\alpha'} \bar{\mu}_\varepsilon^* \bar{\mu}_{\varepsilon'}^* \nu_{\varepsilon\alpha} \nu_{\varepsilon'\alpha'}\right) \ket{\bar{1}0} .
\end{equation}

The same derivation for the initial term $\ket{\bar{1}1}$ gives the mirror expression under $l \leftrightarrow r$,
\begin{equation}
    \left(\Gamma \, e^{-\kappa_{\rm NV}(t_1 - t_m)/2} \sum_{\alpha, \alpha', \varepsilon, \varepsilon'} e^{-i\varepsilon t_m - i\varepsilon' t_m' - i(k_\alpha x_r + k_{\alpha'} x_l)} \, O_\alpha O_{\alpha'} \bar{\mu}_\varepsilon^* \bar{\mu}_{\varepsilon'}^* \nu_{\varepsilon\alpha} \nu_{\varepsilon'\alpha'}\right) \ket{0\bar{1}} .
\end{equation}

The amplitudes of $\ket{\bar{1}0}$ and $\ket{0\bar{1}}$ are then equal provided that, for each $\alpha, \alpha'$,
\begin{equation}
    e^{-i(k_\alpha - k_{\alpha'})(x_l - x_r)} = 1 .
\end{equation}
This is the same condition found above for the two single-excitation sectors to share eigenvalues. When it holds, the heralded state is the target Bell state,
\begin{equation}
    \ket{\psi} \propto \ket{0\bar{1}} + \ket{\bar{1}0} .
\end{equation}

\subsection{Effect of magnon dephasing} \label{sec:magdeph}
In the above, we neglected the effect of pure dephasing by magnons, which amounts to a collapse channel of the form 
\begin{equation}
    L_\alpha^\phi = \sqrt{\kappa_\alpha^\phi} \, m_\alpha^\dagger m_\alpha ,
\end{equation}
with $\kappa_\alpha^\phi$ the pure dephasing rate of mode $\alpha$. It commutes with the magnon numbers and with the NV populations, so every subspace of the table above maps to itself. The sequence of sectors traversed by the protocol is unchanged.

In the non-Hermitian Hamiltonian, this channel adds $-\frac{i}{2} \sum_\alpha \kappa_\alpha^\phi \left(m_\alpha^\dagger m_\alpha\right)^2$. The sectors relevant to the protocol carry at most one magnon, where $\left(m_\alpha^\dagger m_\alpha\right)^2 = m_\alpha^\dagger m_\alpha$. The only change is therefore a renormalized magnon linewidth,
\begin{equation}
    \bar{\eta}_\alpha = \eta_\alpha - i \frac{\kappa_\alpha + \kappa_\alpha^\phi + \kappa_{\rm NV}}{2} .
\end{equation}
The eigenvalue equations keep their form. The eigenvalues $\varepsilon$ and the coefficients $\mu_\varepsilon, \nu_{\varepsilon\alpha}$ only take new values.

Both $L_\alpha^\phi$ and the term it adds to $\widetilde{H}$ act on the magnon alone. They do not depend on $x_j$, nor on which NV is excited. Their action on $\mathcal{H}^1_l$ and on $\mathcal{H}^1_r$ is therefore identical. A dephasing jump at any time projects onto the magnon component of mode $\alpha$,
\begin{equation}
    L_\alpha^\phi \ket{l(\varepsilon)} \propto \nu_{\varepsilon\alpha} \, e^{-ik_\alpha x_l} \ket{0\bar{1}\alpha} , \qquad
    L_\alpha^\phi \ket{r(\varepsilon)} \propto \nu_{\varepsilon\alpha} \, e^{-ik_\alpha x_r} \ket{\bar{1}0\alpha} ,
\end{equation}
with the same coefficient $\nu_{\varepsilon\alpha}$ in both sectors.

The mirror symmetry used in the previous subsection is thus intact. Repeating that derivation with any number of dephasing jumps, at any times, the amplitudes of $\ket{\bar{1}0}$ and $\ket{0\bar{1}}$ stay equal under the same condition $e^{-i(k_\alpha - k_{\alpha'})(x_l - x_r)} = 1$. The heralded state is still $\ket{0\bar{1}} + \ket{\bar{1}0}$, so magnon dephasing leaves the fidelity unchanged. It does shift the eigenvalues $\varepsilon$, so it can change the success probability.

\section{Magnon modes of the wire and their coupling to the NV centers} \label{sec:modes}
In this appendix, we derive the magnon parameters used above. We start from the dispersion of a cylindrical wire, then solve the magnetostatic problem for the stray field of a single magnon. Evaluating that field at the NV positions gives the coupling $g_\alpha$, along with the numerical values quoted in the main text.

\subsection{Dispersion} \label{sec:dispersion}

We model the magnet as a cylindrical wire of radius $d = \SI{40}{\nano\meter}$ and length $L = \SI{10}{\micro\meter}$, with the equilibrium magnetization along the wire axis $\hat{z}$. We consider spin waves with wavevector $k$ along the axis, uniform across the cross-section. Since $\boldsymbol{k} \mathbin{\parallel} \boldsymbol{M}$, these are backward-volume modes. We approximate their dispersion by the corresponding thin-film expression~\cite{Kalinikos_Slavin}, replacing the film thickness by the wire radius,
\begin{equation}
    \omega_k^2 = \left(\omega_a + D_{\rm ex} k^2\right) \left[\omega_a + D_{\rm ex} k^2 + \omega_s \left(\frac{1 - e^{-kd}}{kd}\right)\right]. \label{Seq:dispersion}
\end{equation}
Here, $\omega_a = \gamma B_{\rm ext}$ is the Zeeman frequency of the applied field $B_{\rm ext} = \SI{70}{\milli\tesla}$, with $\gamma = 2\pi \times \SI{28}{\giga\hertz\per\tesla}$ the gyromagnetic ratio. The frequency $\omega_s = \gamma \mu_0 M_s = 2\pi \times \SI{5.28}{\giga\hertz}$ is set by the saturation magnetization $M_s = \SI{150}{\kilo\ampere\per\meter}$. The exchange stiffness is parametrized as $D_{\rm ex} = \omega_s / k_{\rm ex}^2$, where $k_{\rm ex} = 2\pi / \lambda_{\rm ex}$ with $\lambda_{\rm ex} = \SI{110}{\nano\meter}$. These values are representative of yttrium iron garnet. The dispersion has a minimum of $2\pi \times \SI{3.68}{\giga\hertz}$ at a finite wavevector, which sets the magnon gap quoted in the main text. Each mode decays at the rate $\kappa_\alpha = \alpha_G\, \omega_\alpha$, where $\alpha_G = 2 \times 10^{-5}$ is the Gilbert damping. This is a population decay rate. Pure magnon dephasing is a separate channel, treated in Appendix~\ref{sec:magdeph}. At the NV frequency, this gives the linewidth $\kappa \approx 2\pi \times \SI{0.1}{\mega\hertz}$ used in the main text.

This substitution is an estimate rather than an exact treatment. The exact modes of a cylindrical wire require a numerical micromagnetic calculation. We avoid it because the dispersion enters our results only weakly. The only magnetic parameter used below is the coupling $g_\alpha$. It depends on the dispersion through the resonant wavevector $k_{\rm mag}$, and this dependence is weak. A $25\%$ change in $k_{\rm mag}$ changes $g$ by less than $15\%$. A similarly weak dependence was found for a rectangular wire~\cite{Martijn_Entanglement}. We further assume unpinned surface spins, for which a uniform transverse profile is well justified for a nanoscale wire at $\sim 2\pi \times \SI{5}{\giga\hertz}$. Pinned surface spins would give a non-uniform transverse profile, and hence a different coupling. The three-magnon triplet used for detection requires only energy and momentum conservation. Such a triplet exists for a wide range of dispersions.

\subsection{Stray field}

To find the coupling, we need the stray field generated by a magnon. Consider the mode with wavevector $k$. Its transverse magnetization is uniform across the cross-section and circularly polarized, $m_y = im_x \propto e^{ikz}$. In cylindrical components $(\rho, \phi, z)$,
\begin{equation}
    m_\rho = m_0\, e^{i\phi} e^{ikz - i\omega t}, \qquad m_\phi = im_0\, e^{i\phi} e^{ikz - i\omega t}, \label{Seq:magnetization}
\end{equation}
where the magnetization is measured in units of $M_s$, so the amplitude $m_0$ is dimensionless. This profile has no volume charge, $\nabla \cdot \boldsymbol{m} = 0$. The magnetostatic potential $\psi$, defined by the stray field $\boldsymbol{h} = -\nabla\psi$ in units of $M_s$, thus satisfies Laplace's equation both inside and outside the wire,
\begin{equation}
    \nabla^2 \psi = 0,
\end{equation}
subject to continuity of $\psi$ and of the normal magnetic induction at the surface $\rho = d$,
\begin{equation}
    \left(-\partial_\rho \psi + m_\rho\right)^{\rm in}_{\rho = d} = \left(-\partial_\rho \psi\right)^{\rm out}_{\rho = d}. \label{Seq:BC}
\end{equation}
Since the source $m_\rho$ carries a single azimuthal harmonic $e^{i\phi}$, only that component of $\psi$ is excited. Writing $\psi = \psi_0\, e^{i\phi} e^{ikz - i\omega t} f(\rho)$, Laplace's equation reduces to the modified Bessel equation of order one,
\begin{equation}
    \rho^2 f'' + \rho f' - \left(k^2 \rho^2 + 1\right) f = 0. \label{Seq:bessel_eq}
\end{equation}
Imposing regularity at $\rho = 0$, decay as $\rho \rightarrow \infty$, and continuity of $\psi$ across $\rho = d$ gives
\begin{equation}
    \psi = \psi_0\, e^{i\phi} e^{ikz - i\omega t}
    \begin{cases}
        \dfrac{K_1(k\rho)}{K_1(kd)}, & \rho > d, \\[10pt]
        \dfrac{I_1(k\rho)}{I_1(kd)}, & \rho < d.
    \end{cases} \label{Seq:psi_form}
\end{equation}
The amplitude $\psi_0$ follows from the boundary condition Eq.~\eqref{Seq:BC}. Inserting the two branches and using the Wronskian of Bessel functions, $I_1 K_1' - I_1' K_1 = -1/\rho$, we find
\begin{equation}
    m_0 = \psi_0 \left(\frac{I_1'(k\rho)}{I_1(k\rho)} - \frac{K_1'(k\rho)}{K_1(k\rho)}\right)_{\rho = d}
    = \left(\frac{\psi_0}{\rho\, I_1(k\rho) K_1(k\rho)}\right)_{\rho = d}, \label{Seq:psi_amplitude}
\end{equation}
where a prime denotes $\partial_\rho$. Hence $\psi_0 = d\, m_0\, I_1(kd) K_1(kd)$, so that
\begin{equation}
    \psi = d\, m_0\, e^{i\phi} e^{ikz - i\omega t}
    \begin{cases}
        I_1(kd) K_1(k\rho), & \rho > d, \\[6pt]
        K_1(kd) I_1(k\rho), & \rho < d.
    \end{cases} \label{Seq:psi_solution}
\end{equation}

Finally, the zero-point amplitude of the magnetization is fixed by normalizing to a single magnon in the wire volume $V_m = \pi d^2 L$~\cite{OMagCrystal_Jasmin},
\begin{equation}
    m_0 = \sqrt{\frac{\gamma\hbar}{2M_s V_m}} = \sqrt{\frac{\gamma^2 \hbar \mu_0}{2\omega_s V_m}}. \label{Seq:zero_point}
\end{equation}
For our parameters, $m_0 = \num{3.51e-5}$.

\subsection{Coupling to the NV centers}

The NVs sit outside the wire, where the magnetization vanishes and the stray field is $\boldsymbol{B} = \mu_0 M_s \boldsymbol{h}$ with $\boldsymbol{h} = -\nabla\psi$. Each NV couples to it via the Zeeman interaction
\begin{equation}
    \frac{H_Z}{\hbar} = \gamma\, \boldsymbol{B}(\boldsymbol{r}_j) \cdot \boldsymbol{S}^j = \omega_s\, \boldsymbol{h}(\boldsymbol{r}_j) \cdot \boldsymbol{S}^j, \label{Seq:zeeman}
\end{equation}
where $\boldsymbol{S}^j$ is the spin-1 operator of NV $j$ and we take the NV gyromagnetic ratio equal to $\gamma$. We take the NV axis along the wire, so the states $\{\ket{0}, \ket{1}, \ket{\bar{1}}\}$ are eigenstates of $S_z^j$. With the zero-field splitting $D_{\rm NV} = 2\pi \times \SI{2.87}{\giga\hertz}$, the transition frequencies are $\omega_1 = D_{\rm NV} + \gamma B_{\rm ext} = 2\pi \times \SI{4.83}{\giga\hertz}$ and $\omega_{\bar{1}} = D_{\rm NV} - \gamma B_{\rm ext} = 2\pi \times \SI{0.91}{\giga\hertz}$, as quoted in the main text.

The relevant matrix elements are $\braket{1|S_x|0} = 1/\sqrt{2}$ and $\braket{1|S_y|0} = -i/\sqrt{2}$. Keeping the term resonant with the $\ket{0} \leftrightarrow \ket{1}$ transition, the interaction reads 
\begin{equation} 
    \frac{H_{\rm int}}{\hbar} = \sum_{\alpha j} (\omega_s/\sqrt{2}) \left(h_x - ih_y\right)_\alpha \! (\boldsymbol{r}_j)\, m_\alpha P_{01}^j + {\rm h.c.},
\end{equation}
where $\boldsymbol{h}_\alpha$ is the stray field of a single magnon in mode $\alpha$, found above. Converting to cylindrical components, $h_x - ih_y = e^{-i\phi} \left(h_\rho - ih_\phi\right)$. For the outside branch of Eq.~\eqref{Seq:psi_solution},
\begin{equation}
    h_\rho - ih_\phi = -\partial_\rho \psi - \frac{\psi}{\rho} = d\, m_0\, k\, I_1(kd) K_0(k\rho)\, e^{i\phi} e^{ikz}, \label{Seq:h_minus}
\end{equation}
using $K_1'(z) + K_1(z)/z = -K_0(z)$. The azimuthal phases cancel, and the factor $e^{ikz_j}$ is the position-dependent phase $e^{ik_\alpha x_j}$ of Eq.~\eqref{Seq:ham}. Comparing the two expressions, the coupling amplitude is
\begin{equation}
    g_\alpha = \frac{\omega_s}{\sqrt{2}}\, k_\alpha d\, I_1(k_\alpha d)\, K_0(k_\alpha \rho_{\rm NV})\, m_0, \label{Seq:coupling}
\end{equation}
where $\rho_{\rm NV} = \SI{45}{\nano\meter}$ is the radial distance of the NVs from the wire axis, i.e., $\SI{5}{\nano\meter}$ from the surface. At the resonant wavevector $k_{\rm mag} = 2\pi/\SI{200}{\nano\meter}$, this gives $g = 2\pi \times \SI{30}{\kilo\hertz}$, the value used in the main text.

The remaining components of the stray field do not affect the protocol. The longitudinal component $h_z$ couples to $S_z^j$. It modulates the NV frequencies at the magnon frequency of several gigahertz, so it averages out. The counter-rotating combination $h_x + ih_y$ couples to the $\ket{0} \leftrightarrow \ket{\bar{1}}$ transition with a comparable magnitude. That transition lies inside the magnon gap, detuned by more than $2\pi \times \SI{2.7}{\giga\hertz}$ from every magnon mode, so this coupling is strongly off-resonant and we neglect it. This spectral protection is what keeps the computational basis free of magnon-induced decay, as claimed in the main text.

\bibliography{References}

\end{document}